\documentclass{aastex}
\usepackage{spr-astr-addons}

\RequirePackage{color}

\newcommand{\emaila}{giuseppe.altavilla@oabo.inaf.it}

\begin{document}

\title{Supernovae and Gaia}
\slugcomment{Not to appear in Nonlearned J., 45.}
\shorttitle{Gaia \& SNe}
\shortauthors{Altavilla et al.}

\author{Giuseppe Altavilla\altaffilmark{1}} 
\affil{INAF-Osservatorio Astronomico di Bologna}
\and \author{Maria Teresa Botticella\altaffilmark{2}}
\affil{INAF-Osservatorio Astronomico di Padova}
\and \author{Enrico Cappellaro \altaffilmark{2}}
\affil{INAF-Osservatorio Astronomico di Padova}
\and \author{Massimo Turatto \altaffilmark{3}}
\affil{INAF-Osservatorio Astronomico di Trieste}

\email{\emaila}


\begin{abstract}

Despite decades of dedicated efforts there are still basic questions to answer 
with  regard to Supernova progenitor systems and  explosion mechanisms. In 
particular, in the  last years a number of exceptionally bright objects and 
extremely faint events have  demonstrated an unexpected large  Supernova variety.

The large number of Supernovae candidates at different redshifts provided by 
the  next  generation surveys, from ground and space,  will allow  to reach
a better insight of the Supernova events in all their flavours.  In particular
it will be the possible to assess the systematics of type Ia Supernovae 
as distance indicator at any redshift.

The Gaia astrometric mission is expected to discover a huge number of transient 
events, including Supernovae, which will be immediately disseminated to the 
astronomical community by  a transients alert  system for a suitable  follow up.
\end{abstract}

\section{SNe}\label{sec:SNe}
Supernovae (SNe) are dramatic and violent end-points of stellar evolution, and lie at the heart of some of the
most important problems of modern astronomy since their huge  luminosities were first estimated and a clear distinction between classical
novae and SNe was made \citep{baade1934b}.

{\bf SNe and stellar evolution}\\
SNe provide natural laboratories for studying the physics of  hydrodynamic, nuclear processes in extreme conditions and are  involved in the
formation of neutron stars, black holes, and gamma-ray bursts (GRBs).
Furthermore,  SNe are sources of gravitational waves and of neutrino emission, and candidate sites for high-energy cosmic ray acceleration. 
Thus, it is of broad astrophysical importance to understand the physical conditions for these spectacular explosions.
\\
We recognize two physically defined classes: core collapse-induced explosion of short-lived massive stars (CC SNe) and thermonuclear explosions of long-lived low mass stars (SNe Ia).
In recent years deeper and more frequent searches for transient events in the local and distant Universe  as well the identification of  SN progenitor stars on  pre-discovery images from Hubble and Spitzer Space telescopes  have provided us with important information on the evolution of  massive stars and binary systems \citep{Smartt2009}. 
{\it However,  a  growing number of peculiar events  suggests that the standard scenario of stellar evolution and explosion physics
may not be complete and demands further physical understanding and theoretical explanation. }

{\bf SNe and dark energy}\\
Type Ia SNe are standard candles due to their significant intrinsic brightness and homogeneity.  They
have provided the first evidence for an accelerated expansion of the Universe and   remain one of the more promising probes of  the nature and amount of dark energy,
\citep{riess1998,perlmutter1999,tonry2003,astier2006}. 
CC SNe display a huge range in their physical characteristics, including kinetic energy, radiated energy and the amount of radioactive elements  created during the explosion.  
However, plateau  CC SNe  (SNe IIP)  that undergo a long phase of constant luminosity in their photometric evolution are potential standardizable candles,  although they are not as luminous and uniform as
SNe Ia \citep{Nugent2006}.
{\it Understanding the mechanism that is responsible for the accelerating expansion of the Universe is one of the crucial next steps for cosmology and requires  both new searches for type Ia  and type IIP SNe  to populate the Hubble diagram and the knowledge of their progenitors and explosion mechanisms. }

{\bf SNe and galaxy evolution}\\
SNe  play  also a major role in driving the chemical, kinetic, and dynamical evolution of galaxies. 
They are the main producers of heavy elements and are fundamental for understanding global abundances and abundance patterns in galaxy clusters and in the intergalactic medium \citep{Matteucci1986}. 
SNe Ia are important contributors of iron and release it to the interstellar medium over longer timescales, compared to the $\alpha$ elements produced by CC SNe.
Therefore the $\alpha$ to iron abundance ratio can be used to evaluate the formation timescales of stellar systems.  
Moreover the SN metal-rich ejecta are believed to be a potentially important site of cosmic dust formation. 
The energy release from SNe can initiate episodes of star formation (SF), impact the evolution of gas flow and contribute to the feedback processes in galaxies \citep{Ceverino2009}.
{\it A complete and coherent picture of the formation and evolution of the galaxies is a fundamental objective of modern astronomy  and SNe are an important ingredient to model  the chemical enrichment of the galaxies and the effect of  energy/momentum feedback on galaxy formation.}

{\bf SNe and SFR}\\
The cosmic evolution of SN rates is an alternative and complementary probe of the SF history (SFH) in the galaxies. 
Due to the short lifetime of progenitor stars the CC SN rate is directly proportional to the current SFR while type Ia SN rate reflects the convolution of the SFH with the distribution of the delay times between progenitor star formation and explosion \citep{Heger2003,Greggio2005}. 
Poor statistics is a major limiting factor for using the CC SN rate as a tracer of the SFR. At low redshift the difficulty is in sampling large enough volumes of the local Universe to ensure significant
statistics \cite[e.g.][]{Botticella2012}, while  at high redshift  the difficulty  lies in detecting and typing complete samples of intrinsically faint SNe \citep{botticella2008,Bazin2009,li2011b}.  Moreover it is expected that a fraction of CC SNe are missed by optical searches, since they are embedded in dusty spiral arms or galaxy nuclei. This fraction may change with redshift, if the amount of dust in galaxies evolves with time.  
{\it Progress in using CC SN rates as SFR tracers requires accurate measurements of rates at various cosmic epochs and in different environments. 
Furthermore it requires a comparison  with other SFR diagnostics  in a wide range of redshift to investigate the causes of the observed discrepancy. }
\\

Despite the importance of SNe in several astronomical topic  many fundamental questions  remain with regard these exciting phenomena.

\subsection{CC SNe}
Stellar evolution theory predicts that all stars more massive than $8-10 M_{\odot} $ complete eso-energetic nuclear burnings and develop an iron core that cannot be supported by any further nuclear fusion
reaction, or by electron degenerate pressure. The collapse of the iron core results in the formation of a compact object, a neutron star or a black hole, accompanied by the high-velocity ejection of a large fraction of the progenitor mass.
Despite this basic understanding, the picture of massive star  death emerging from  recent discovery is far to be clear. 

{\bf Which massive star  in which evolutionary stage produce which CC SN type?}\\
Different types of CC SNe have been identified on the basis of their spectroscopic and photometric properties: type II SNe show presence of H in the spectra, (type IIP SNe exhibit a long plateau in their light curve while type IIL show a linear decline rate), type Ib SNe lack H but show He, type  Ic  SNe lack both H and He, type IIn SNe exhibit a narrow $H\alpha$ emission line and present  other strong evidences of interaction of the ejecta  with circumstellar  material (CSM) already at early phases of their evolution.
This sequence has been explained on the basis of the progenitor mass loss history   with the most massive stars loosing the largest fraction of their initial mass \citep{Heger2003}:  type IIP progenitor stars are red super giants (RSGs) that explode with most of their  H envelope present, type IIL progenitors had extensive mass loss but preserved  a small  H envelope,  type Ib/c progenitors are Wolf-Rayet stars that had lost their H and He envelopes, type IIn progenitor have suffered a major mass loss episode just before the explosion.
Several  CC SNe   with characteristics intermediate between these canonical types  have been  recently discovered   suggesting a smooth  transition and requiring the introduction of hybrid classes (for example type IIb SNe  and  Ibn SNe). 
The progenitor stars of SNe IIP  are the best constrained from direct detection since these SNe are the most frequent in the nearby Universe. Their mass function  seems  to suggest a lack of high mass progenitors  ($ >17 M_{\odot}$) and  raises the intriguing possibility that  more massive stars collapse directly to a black-hole and produce either faint or failed SNe  \citep{Smartt2009b}. 
The progenitors of SNe Ib/c have so far eluded discovery and several observational evidences have suggested an alternative scenario  where a significant fraction of their  progenitors are stars of much lower initial mass in close binary systems stripped of their envelopes  by the interaction with the companion \citep{Podsiadlowski2004,Eldridge2011}. 
Recent observations  have suggested the possibility that stars explode as CC SNe shortly after  the luminous blue variable (LBV)  phase characterized by  sporadic variability and occasionally giant  outbursts  with a huge mass loss in a single short event  \citep{Humphreys1994}. 
In particular some SNe (2006jc, 2006gy,  \citealt{Pastorello2008,Smith2007, Agnoletto2009}) showed evidence that their progenitors  had properties consistent with LBV and in one case a pre-explosion outburst was seen.

{ \bf Which physical parameters  drive  the CC SN diversity?}\\
The variety of the observational properties shown by CC SNe and the relative numbers of  different  types  reflects the large range of stellar types seen in the upper region of the Hertzsprung Russel Diagram above $ 8 M_{\odot}$  and the different physical conditions (radius, density profile and surrounding CSM) before the explosion \citep{Hirschi2010}.
The simple  scheme where only the mass loss drives the evolution of massive stars can not explain the surprising  diversity  in explosion energies and   it is essential to investigate  the effect of other  key factors  (metallicity, rotation rate, magnetic fields, binary membership)   \citep{Heger2005, Eldridge2004, Hirschi2005, Yoon2005}.

A major puzzle is due to the discovery of a number of SNe at the extrema of the luminosity function and requires  further physical understanding. 
A subset of  sub-luminous type II SNe shows both a very faint radioactive tail in the light curve and a low
expansion velocity suggesting low energy explosions  and a  possible black hole remnant, giving rise to significant fallback  \citep{Zampieri2003,Pastorello2004}. 

Different types of  extremely luminous  CC SNe have been recently discovered in dwarf,  faint host galaxies \citep{Botticella2010,quimby2007,pastorello2010} with an enormous luminosity at maximum  that can not be powered by radioactive
 $^{56}$Ni.
Several scenarios have been proposed  to explain these events:  an explosion driven by black hole accretion \citep{Utrobin2010}, the spin down of a rapidly rotating young magnetar \citep{Kasen2010, Woosley2010}, interaction of the SN ejecta with a dense CSM \citep{Blinnikov2010,Chevalier2011} or pulsational pair instability in which collisions between high velocity shells are the source of multiple, bright optical transients \citep{Woosley2007}.

{\bf What are the alternative explosion mechanisms producing a CC SN? }\\
To explain the full range of explosion parameters of CC SNe  some theoretical studies suggested two alternative explosion mechanisms with respect to the collapse of the iron core:  electron capture   and pair instability. 
Electron capture SNe (EC SNe) occur when a star in the mass range of $\sim 8-10 M_{\odot}$ forms an electron degenerate O-Ne-Mg core and the core may collapse before Ne ignition  \citep{Nomoto1984,Ritossa1999,Siess2007,Poelarends2008}. 
The lower and maximum initial mass for EC SNe  is determined by  the competing effects of  core growth and mass loss during the late evolutionary stages of super-AGB stars and are quite uncertain  ($\sim 9- 9.25 M_{\odot}$, \citealt{Siess2007,Poelarends2008}).
The pair  instability SNe  (PI SNe) occur  when the high temperatures  in a massive core (He core  $\geq 40 M_{\odot}$ and initial mass $\geq 100M_{\odot}$)  induces  electron-positron  pair production reducing the radiation pressure and hence leading to a rapid contraction followed by a thermonuclear explosion.
A similar mechanism  is pulsational pair instability  in which a massive core  undergoes  interior instability  again due to electron-positron pair production ejecting  many solar masses  in a series of giant pulses  \citep{Woosley2007}. 
At the moment  there are few observational probes of PI SNe.

{\bf Which type of massive stars produce black holes? }\\
The origin of stellar mass black holes is not yet understood:  some stars ($25 -40 M_{\odot} $) are expected to produce  weak explosions  with a black hole formed by fallback, some others ($>40 M_{\odot} $) collapse into a black hole directly without any optical signature (failed SNe).

The failed SNe have been invoked by theorists because of the difficulties of producing an explosion in analytical models   \cite[e.g.] {Heger2003, Eldridge2004} but  there are remarkably few observational probes of them, e.g. the lack of high mass progenitors may suggest there is a population of black-hole forming SNe which so far have eluded discovery.
The detection of sudden neutrino bursts with given duration and energy (to discriminate other events, such as ordinary SNe, neutron stars formations and so on), may be pontentially the signature of failed SNe \citep{yang2011,sumiyoshi2010}.
Also gamma-ray astronomy \citep{MacFadyen1999} and gravitational-wave astronomy may be  sensitive probes of failed SNe \citep{kotake2010}.

{\bf  What is the nature of SN impostors?}\\
The luminosity gap separating novae and sub-luminous SNe is recently being populated by a number of  low energy events  that have been classified as  SNe (SN impostors) and show a significant diversity  ranging from giant eruptions of  LBVs  to  very faint transients of unknown physical origin \citep{Smith2011}.
The  overlapping of bolometric luminosity and kinetic energies between SN impostors and faint CC SNe requires to investigate in more detail the observational differences between explosive  and eruptive transients.
At the moment the only certain test to distinguish stellar eruptions from genuine CC SN explosions is to observe the outcome of the transient event: the surviving progenitor star in the case of eruptive event or a compact SN remnant  in the case of  CC SN explosion \citep{Chevalier2005}.
 We also need  to understand the diversity of these transients  that may reflect different eruption or explosion mechanisms  (luminous red novae, stellar mergers in low mass
systems, erupting OH/IR stars, weak EC SNe from super-AGB stars) \citep{Kulkarni2007,Prieto2008,Botticella2009,Kochanek2011}.

{ \bf What distinguishes a GRB progenitor from that of hypernovae?}\\
The energetically most extreme SNe  Ib/Ic (hypernovae) have been associated to long GRBs \citep{Woosley2006,Campana2008,Modjaz2011}.
GRB-SNe are strongly aspherical explosions characterized by a very luminous maximum light, a huge expansion velocity and a very high kinetic energy  (about 10 larger than observed in standard CC SNe) \citep{Maeda2008}. 
 Their progenitors are  massive stars ($30 - 40 M_{\odot}$)  in low metallicity star forming environments \citep{Fruchter2006,Modjaz2008} with high rotation rate  \citep{Yoon2005} that  lost  all H envelope.  
To date  only six clear cases of hypernovae have been discovered \footnote{SN 1998bw \& GRB 980425
(z=0.0085) \citep{Galama1998}; SN 2003dh \& GRB 030329 (z=0.17) \citep{Stanek2003,Hjorth2003,Matheson2003}; SN 2003lw \& GRB 031202 (z=0.1) \citep{Malesani2004}; SN 2006aj \& GRB060218 (z=0.033) \citep{Campana2006,Pian2006}; SN 2008hw \& GRB081007 (z=0.53) \citep{Dellavalle2008}, SN2010bh \& GRB100316D at z = 0.0593 \citep{Chornock2010,Starling2011}} plus another three cases which show possible SN signatures in the spectra of the GRB afterglows.
The existence of SNe Ic/Ib without observed GRBs, as well as that of GRBs without SN signatures, raises the question of what distinguishes
a long GRB progenitor from that of  a hypernova and an ordinary SN Ic/Ib \citep{dellavalle2006}.

\subsection{SNe Ia}
There is general consensus that  SNe Ia are  thermonuclear explosions of a carbon and oxygen white dwarf  (CO WD) near the Chandrasekhar mass ($1.4 M_{\odot}$) in a binary system. 
The use of SNe Ia as standard candles  is based on the assumption that all SNe Ia have similar progenitors  and are highly homogeneous but  the nature of
the companion star and the details of the explosion mechanism are still debated.

{\bf What  is the progenitor star  and explosion mechanism of SNe Ia?}\\
The most widely favoured progenitor scenarios for SNe Ia are  the single degenerate  scenario (SD) in which a  WD, accreting from a  non degenerate companion (a main sequence star, a subgiant, a red giant or a helium star), grows in mass until it reaches the  Chandrasekhar mass and explodes in a thermonuclear runaway, or double degenerate scenario (DD), in which a close double WD system merges after orbital shrinking due to the emission of gravitational waves \citep{Hillebrandt2000,maoz2011b}.
The time elapsed from the birth of the binary system to the SN explosion (delay-time) spans from tens of million years to ten billion years  \citep{Greggio2005}.
A thermonuclear explosion may occur before the WD reaches the Chandrasekhar mass  (sub-Chandrasekhar models) \cite[and reference therein]{Fink2011} or when the WD  exceed the Chandrasekhar mass (super-Chandrasekhar models)  in both SD and DD scenario \citep{Hachisu2011}.
The observed features of  standard SNe Ia are better explained by the Chandrasekhar mass although no conclusive evidence for a specific scenario has so far been found.
The recent discovery of variable circumstellar absorption lines in the SNIa 2006X \citep{Patat2007} and  the first direct detection
of the precursor of a SN Ia (an X-Ray source at the position of SN 2007on,  \citealt{Voss2008}) seem to support the SD scenario.
However \cite{roelofs2008}  did not detect  any X-ray  source in images taken six weeks after SN 2007on optical  maximum  and found an offset between the SN and the measured X-ray source position of $1.15\pm0.27$".  Only future observations can shed light on the proposed connection between the X-ray source and the progenitor of SN 2007on, and thus on the most  plausible scenario for this SN Ia progenitor.
The analysis of the recent SN 2011fe in M101 seems to rule out the presence
of a red giant and helium-star donors in pre-explosion images \citep{li2011c}, of a circumstellar
wind from a giant donor \citep{horesh2012} and  of shocks from ejecta hitting a companion, by analysing very early
optical and UV data \citep{nugent2011,brown2011,bloom2012}. 

The explosion mechanism of standard SNe Ia  is also largely unknown (see \citealt{Hillebrandt2000} for a review). Key issues include the initiation of the thermonuclear runaway, its initial ignition
point and subsequent evolution.
In the Chandrasekhar mass scenario, a WD is thought to ignite near the center. At first the flame propagates subsonically as a deflagration and in a second phase a detonation triggers that propagates supersonically.
The mechanism that leads to the formation of the detonation is still an open question: detonation transition by turbulence (e.g. \citealt{Ropke2007}) gravitationally confined detonation (e.g.  \citealt{Jordan2008,Meakin2009})
and pulsating reverse detonation \citep{Bravo2009}.



{\bf Why does their brightness correlate with light curve shape, color, stellar population age, metallicity?}\\
Thanks to the uniformity of  their peak luminosities SNe Ia  can be used as cosmological standard candles.
A phenomenological correlation between their absolute magnitude and  the shape of their light curves (with brighter
objects having a slower rate of decline) has been provided to calibrate their distances \citep{Phillips1993,Phillips1999}. 
Only recently  models were able to reproduce this calibration  in terms of the underlying physics \citep{Kasen2009}. 
However, this one-parameter description does not completely account for the variety of SNe Ia.  The relation between color and light curve shape  (brighter SNe Ia are both bluer and have wider
light curves than their fainter counterparts) has been exploited  to  provide a more
accurate luminosity calibration.  However, the intrinsic color variations of SNe Ia
have not yet been fully understood. 
Several suggestions have been
made for additional parameters:  metallicity \citep{Gallagher2005,Hoflich2010}, high-velocity spectral features, spectral flux ratios \citep{Bailey2009}, and the mass and/or the morphological type
of the host galaxy \citep{Kelly2010,Sullivan2010}. 
An interesting possibility for the origin of the diverse
properties of SNe Ia was recently suggested by \cite{Kasen2009} theoretically and by \cite{Maeda2011} observationally, namely an asymmetry in
the SN explosion combined with the observer viewing angle.

{\bf What  is the origin  of SN Ia diversity?}\\
The standard SNe Ia  used as cosmological distance indicators make up about 70\%  of the observed SNe Ia \citep{li2011a}. 
With the advent of the new wide-area transient surveys,  spectroscopic and photometric peculiarities have been noted in SNe Ia with increasing frequency   and new subclasses of SNe Ia  \citep{Benetti2005}  have been introduced:  sub-luminous (e.g. SN 1991bg), normal, bright (e.g. SN 1991T) and super-luminous events (e.g. SN 2007if).
Whether they form distinct physical groups from normal SNe Ia, with different progenitors and explosion models, or whether they lie at the extreme end of a continuous distribution  is unclear. 
Recent observations of sub-luminous and super-luminous SNe Ia\footnote{SN 2003fg SN 2006gz, SN 2007if, and
SN 2009dc}  suggest that their WD progenitors might  have sub-Chandrasekhar and super-Chandrasekhar mass, respectively \citep{Hicken2007,Scalzo2010,Taubenberger2011}.
Various  scenarios, for example a standard explosion with strong deviations from spherical symmetry  of a Chandrasekhar mass WD \citep{Hillebrandt2007}, or alternative energy sources other than radioactivity have been proposed to explain the super-luminous  SNe. 
It is also established that sub-luminous SNe Ia preferably occur in non-star-forming host galaxies with large stellar
masses, such as ellipticals \citep{Neill2009}  while super-luminous SNe Ia  occur in   relatively metal poor host galaxies \citep{Taubenberger2011}
Understanding super-luminous and sub-luminous SNe Ia has important implications for cosmology and  may challenge the paradigm of SN Ia progenitors. 

Modern SN searches are revealing ever more extreme examples of faint SNe Ia with  lower  kinetic energies, the nature of which   is controversial  with possible models ranging from the direct
collapse of a massive star to a black hole  to the pure deflagration of a  WD (e.g., \citealt{Foley2010})

{\bf  Are there two different progenitor channels?}\\
A distinct population of SN Ia progenitors (called prompt) with short delay time  (few hundred million years) has been proposed to explain the high SN Ia rate in blue  star-forming galaxies
and in radio-loud galaxies  \citep{Mannucci2005,Greggio2008}.
 Tardy SNe Ia arise in old population of at least several Gyr  in red, passively evolving
ellipticals \citep{Sullivan2006}.
The prompt SNe Ia seem to have broader light curves, to be on average brighter and  to show  fewer intermediate-mass elements than their tardy counterparts. 
The ratio of prompt to tardy events  is expected to change  with redshift,
tracking increasing SFR, with a resulting evolution in average light curve width, SN
Ia brightness, and spectral feature strength.
The current measurements of type Ia SN rates can not constrain the progenitor models for the uncertainties in the SF history and in SN rate measurements.
Progress in investigating the nature of type Ia SN progenitors requires to link SN rates and parent
stellar populations with measurements in star forming and in passively evolving galaxies over a wide
range of redshifts   from $z =0.01$  to $z \sim 4$ \citep{Greggio2010}.

\section{SN surveys \& Gaia}\label{sec:SNsurveysandGaia}

\subsection{Historical SN searches}\label{sec:historicalSNsearches}

The first  systematic  supernova search was undertaken in 1934  by Zwicky 
using  a modest 3-1/4 inch  lens camera.
In 1936 he continued the search
using  photographic plates obtained with the 18 inch Schmidt
telescope at Palomar Observatory  \citep{baade1938}. 
This program was followed by the Palomar Supernova Search 
\citep{humason1960}, performed  
with the 18 and 48 inch Schmidt telescopes, 
which lasted from 1959 till 1974 and discovered
a total of  $\simeq$190 SNe
\citep{zwicky1938,zwicky1942,zwicky1964}.
\\
Several  SN searches followed. We list here a few examples:
the Asiago SN search, that covered the period 1959-1990,
and was performed with the 40/50 cm Schmidt until 1967 and  with the  67/92 cm 
Schmidt afterward. This search produced $\simeq$50 SNe, most of them discovered by 
L.~Rosino on photographic plates \citep{cappellaro1993,cappellaro1997}.
\\
The Crimean SN search operated in the period  1961-1991, using the  40 cm 
astrograph of the Sternberg Institute in Crimea,   
and announced $\simeq$40 SNe  
\citep{tsvetkov1983}.
\\
Reverend Evans’ visual search produced 24 SNe in the period  1980-1988 using 25-41 cm 
telescopes  \citep{vandenbergh1987,evans1989}.
Evans' SNe were usually bright and nearby objects caught soon after explosion
and hence interesting targets for detailed follow-up campaigns.
\\ 
The Observatoire de la C\^ote d’Azur search was performed analyzing photographic
plates acquired for other programmes using the 90/152 cm Schmidt in the period 
1987-1994 
and produced  $\simeq$68 SNe \citep{pollas1994}.
\\
The Cal\'an Tololo search  operated in the period  1989-1995 using the  
60/90 cm  Schmidt. The survey discovered  $\simeq$54 SNe  (on photographic  
plates)  between 1990 and 1993 \citep{hamuy1993}.  This was the first survey 
at redshifts suitable for starting cosmological studies  ($0.01<z <0.1$).

\subsection{Modern and future SN searches}

\begin{figure}[!t]
\includegraphics[angle=0,scale=.41]{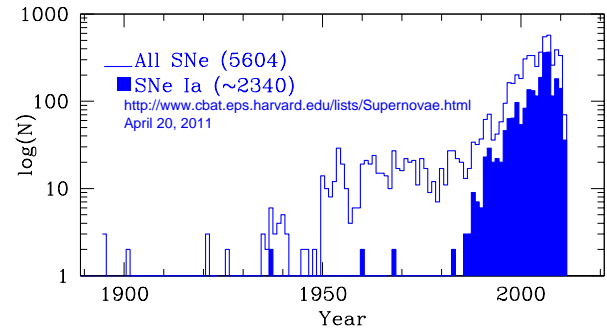}
\caption{SN discovery rate with time. The filled histogram refers to type Ia SNe
} 
\label{fig:sndiscoveryrate1}
\end{figure}

\begin{figure*}[!t]
\includegraphics[angle=0,scale=.41]{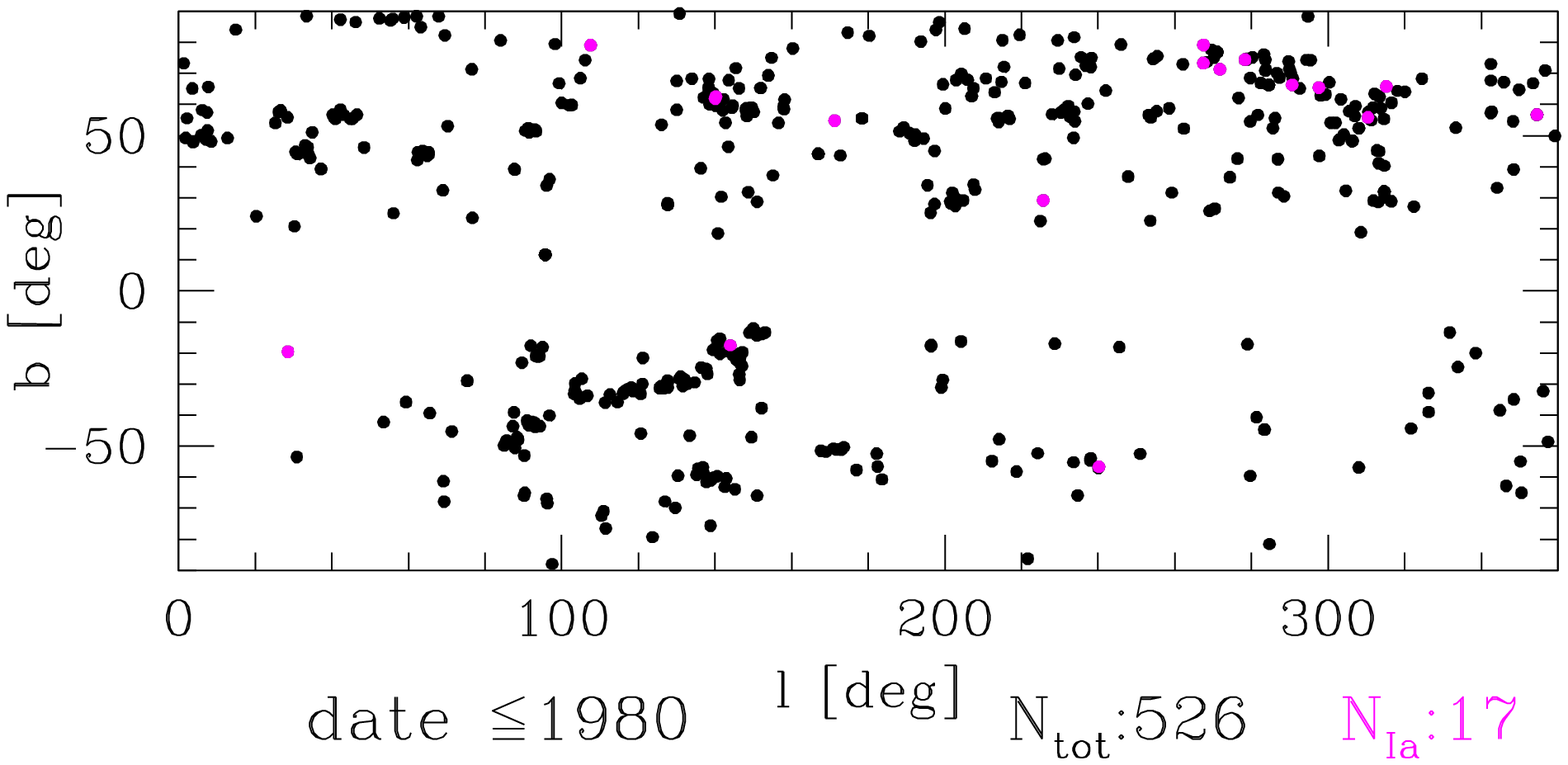}
\includegraphics[angle=0,scale=.41]{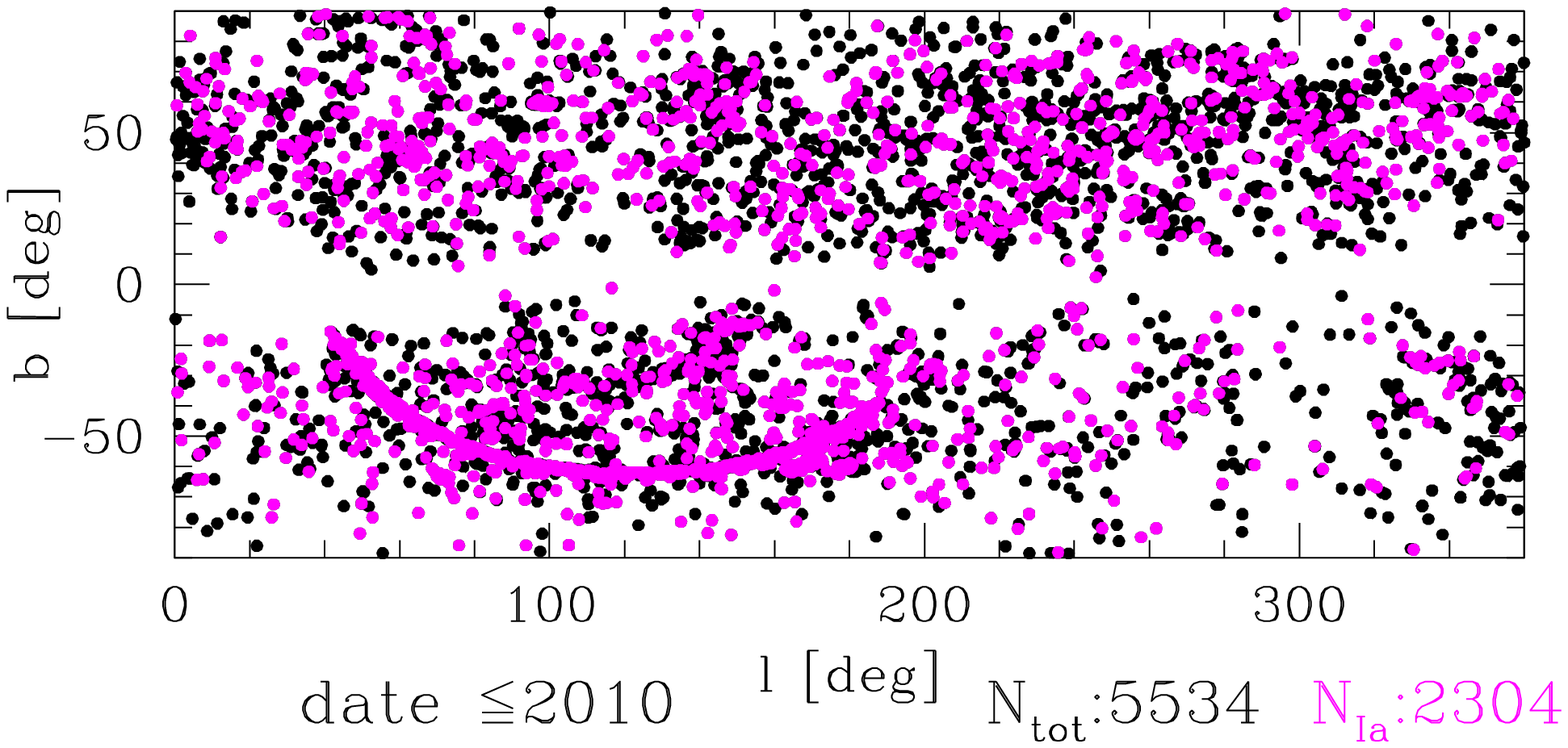}
\caption{Sky distribution  (galactic coordinates) of the SNe  discovered up to 1980 (left panel)
and  up to 2010 (right panel). 
Magenta dots: type Ia SNe; black dots: Core Collapse and unknown SNe. 
Data from http://www.cbat.eps.harvard.edu/lists/Supernovae.html. 
The Milky Way is clearly visible as an empty strip. 
The {\it arc} clearly visible in the
right panel is made up by  the SDSS supernovae 
} 
\label{fig:sndiscoveryrate2}
\end{figure*}

\begin{figure}[!t]
\includegraphics[angle=0,scale=.41]{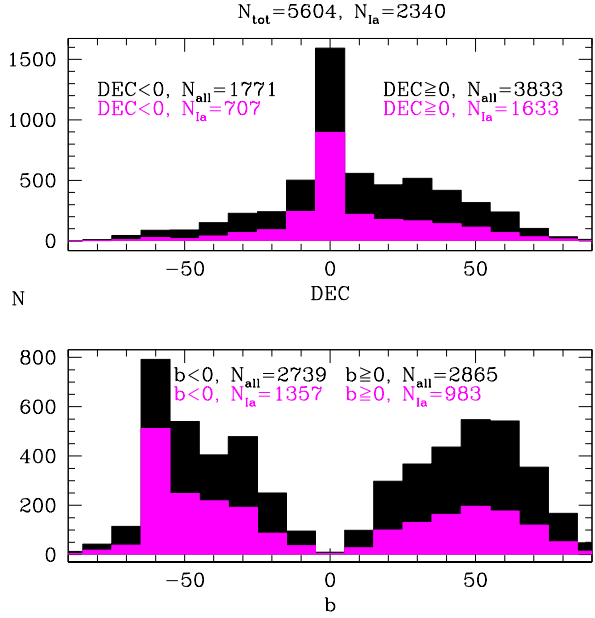}
\caption{SN distribution in the sky (upper panel: equatorial coordinates; 
lower panel galactic coordinates)
} 
\label{fig:sndistribution}
\end{figure}

In the last two  decades of the $20^{th}$ century the advent of CCDs 
revolutionized 
astronomy and visual or photographic observations were 
substituted by digital images. Mechanical blinking machines 
and visual inspections   were replaced by computer
analysis.
The improved CCD and computer technology, the increasing size of telescopes, 
the advent of  larger detectors or CCD mosaics,
the development of automated/robotic telescopes and more  sophisticated image 
analysis software boosted the productivity of SN surveys of the '90s.
Fig.~\ref{fig:sndiscoveryrate1} graphically shows the evolution of the SN 
discovery rate with time and the steep increase that started
at the end of last century.
The same figure shows that a bit less than one half 
of the total  are type Ia SNe. 
Fig.~\ref{fig:sndiscoveryrate2} shows that the number of  SNe discovered
from 1980 to 2010 increased by a factor $\simeq10$ with respect to all SN 
discovered before 1980\footnote{to be noted that our estimate is based on
the SN announced on CBETS and listed in http://www.cbat.eps.harvard.edu/lists/Supernovae.html. 
A not negligible number of SNe discovered
by modern searches are  not present in that list.}. 
We report here a not-exhaustive list 
of modern supernova surveys  to  illustrate this boom:
\begin{itemize}
\item the High-Z SN search \citep{schmidt1998,riess1998,tonry2003};
\item the Supernova Cosmology Project (SCP; \citealt{perlmutter1999,knop2003});
\item the Lick Observatory and the Tenagra Observatory Supernova Searches (LOSS/LOTOSS) 
\citep{filippenko2001,schwartz2000,leaman2011,li2011a,li2011b,maoz2011a};
\item the Perth Automated SN Search \citep{williams1998};
\item the Mount Stromlo Abell Cluster Supernova Search (MSACSS; \citealt{reiss1998});
\item the Wise Observatory Optical Transients Search (WOOTS; \citealt{galyam1998});
\item the CfA Supernova search \citep{riess1999};
\item the Seoul National University SN Search (SNUSS; \citealt{lee1999});
\item the European SN Cosmology Consortium (ESCC; \citealt{hardin1999});
\item the QUEST survey \citep{Schaefer1999,schaefer2000};
\item the Nearby Galaxies SN Search (NGSS; \citealt{strolger2000});
\item the EROS Nearby SN Search (EROSNSS; \citealt{hardin2000});
\item the Beijing Astronomical Observatory SN Survey (BAOSS; \citealt{qiu2001});
\item the All-sky Southern Supernova Search \citep{salvo2003};
\item the UK Nova/Supernova Patrol \citep{hurst2003}; 
\item the Super Livermore Optical Transient Imaging System (Super-LOTIS; \citealt{perezramirez2004});
\item GOODS HST SN Search \citep{strolger2004,riess2007};
\item The Carnegie Supernova Project (CSP; \citealt{hamuy2006,phillips2007,folatelli2010});
\item the Nearby Supernova Factory 
\citep{copin2006};
\item the CFHT Supernova Legacy Survey (SNLS; \citealt{astier2006,conley2011});
\item SkyMapper \citep{keller2007};
\item the ESSENCE Supernova Survey \citep{miknaitis2007,wood-vasey2007,foley2009};
\item the Southern inTermediate Redshift ESO Supernova Search (STRESS; \citealt{botticella2008});
\item SDSS \citep{kessler2009};
\item The CHilean Automatic Supernova sEarch (CHASE; \citealt{pignata2009});
\item the Palomar Transient Factory (PTF; \citealt{law2009,rau2009});
\item the DES Supernova Survey (a dedicated Supernova Survey within
the Dark Energy Survey, DES \citep{sako2011};
\item the Catalina Real-Time Transient Survey (CRTS, \citealt{djorgovski2011});
\end{itemize}
The main characteristic distinguishing the surveys listed above
is  the targeted redshift range:
surveys such as the High-Z SN search, SCP, ESSENCE, CFHT SNLS, GOODS, DES,
are focused on high $z$ SNe while other survey look for closer
events.
Another distinct difference is  the observational strategy:
galaxy-targeted (usual local) SN surveys, monitoring a selected sample
of galaxies or galaxy clusters, such as the LOSS search, 
and  wide field non targeted and unbiased  searches such as SDSS.
Also some searches are not focused on SN only (as the SDSS, PTF
 or the CRTS),  nevertheless they represent a relevant contribution to the SN
discovery rate (see Fig.~\ref{fig:sndiscoveryrate2} upper-right panel).
\\
Fig.~\ref{fig:sndistribution} shows the SN distribution in the sky.
As expected from historical reasons, the northern sky is more populated 
than the  southern, even if  the  SN distribution is quite homogeneous 
in galactic  coordinates, with about one half of the SN  discovered  above 
and below the  galactic plane (the Galaxy prevents the  detection of SNe 
behind the disk,  that is in fact well visible as an empty  strip in 
Fig.~\ref{fig:sndiscoveryrate2}).
\\
In the course of the years the SN searches explored deeper and deeper redshifts.
Each redshift range presents different observational difficulties:
low redshift surveys  ($z<0.1$)  need wide  but shallow field 
observations while   pencil beam  but deep imaging is enough for high redshift 
($z> 0.5$) surveys.
Intermediate redshift searches ($0.1 \lesssim z\lesssim0.5$) 
require wide field and 
deep observations, that have been problematic for a long time.
For this reason for some time Hubble diagrams had a gap  at the  intermediate
 $z$. Nevertheless filling such interval was important for several reasons:
to reduce the statistical uncertainties 
and  to  help detecting  systematic errors such as a possible 
evolution of the supernovae characteristics.
\\
The total number of SN announced in the last years is of the order of a few 
hundred (see Fig.~\ref{fig:sndiscoveryrate1}) 
but several surveys are being planned or have just started that are
expected to significantly increase the discovery rate of SNe
producing (each of them) thousands of detection per years.
Some of the most promising  next generation surveys 
  are  the following:
\begin{itemize}

\item the Panoramic Survey Telescope \& Rapid Response System 
(PAN-STARRS\footnote{\tt http://pan-starrs.ifa.hawaii.edu/public/home.html}),
 an array of  
$4\times1.8$~m telescopes monitoring $\simeq30000$ square degrees in the  northern 
hemisphere (Hawaii). 
This project is expected to detect $10^7$ transients  in total (SN, GRB etc)
and in particular $10^4-10^5$ SNe of all types per year (5000 SNe Ia per year at 
$0<z<1$)\footnote{\tt http://www.astro.caltech.edu/$\sim$avishay/zwicky1/ppts/-\\Price.pdf}. 
The prototype single-mirror telescope PS1\footnote{\tt http://ps1sc.org/transients/} is
already operational on Mount Haleakala, Maui, Hawaii \citep{chomiuk2011}.

\item the Large Synoptic Survey Telescope (LSST\footnote{\tt http://www.lsst.org/lsst/}),
a 8.4~m telescope  monitoring  $\simeq 20000$ square degrees 
in the southern hemisphere (Cerro Pach\'on, Chile).
First light is foreseen in 2018,
scientific observations will start  in 2019 and 
the survey  will be fully operational by 2020.
LSST is expected to detect $\simeq2700$ SNe per night (up to $z\simeq1.2$),
30000 SNe Ia per year at $z<0.3$\footnote{\tt http://www.lsst.org/files/docs/aas/2006/WangL.pdf} and 250000 SNe Ia per year at an average redshift of $<z>\simeq0.45$, 
but with the most distant SN up to $z\simeq0.7$ or even $z\simeq1.4$ with extended exposure time\footnote{\tt http://www.lsst.org/lsst/science/scientist\_transient}. 
Multiband (six colors) photometry will be provided and a fiber spectrograph
can be attached to LSST, but  dedicated spectroscopic follow up is required.
\end{itemize}

These complementary surveys (one in 
the northern and the other in southern sky) will
 provide homogeneous unbiased SN samples order of magnitudes larger
than the current ones. Multicolour light curves will be secured
for each observed object.
The bottle-neck of such large surveys will be the rejection of false positive
detections, the confirmation of the SN candidates and their follow-up.
So far the astronomical community managed to provide 
substantial spectroscopic follow-up for relatively bright  discoveries, 
but did not provide significant help for slightly fainter 
(and much more numerous) events \citep{galyam2011}.
\\
We  note the importance of  untargeted and unbiased 
SN surveys. In the past  supernovae searches
were usually  focused  on large galaxies with high star formation rate 
to maximize the probability of SN detections.
In this case the results were affected by selection effects.
The modern unbiased surveys will  allow us to detect SNe in faint 
dwarf galaxies, hostless {\it orphans} SNe and even new SN classes.
Many supernovae with peculiar new properties, such
as ultra bright SNe, have been  already found 
(e.g. \citealt{quimby2007,gezari2009,pastorello2010,chomiuk2011}).
We expect that the untargeted surveys to
produce an even larger richness in the SN variety that, in turn,
will push for  new explosion mechanism models.

In this context, a significant contribution will 
be offered by the ESA space mission Gaia.

\section{Gaia contribution to SN discovery}

\begin{figure}[!t]
\includegraphics[angle=-90,scale=.30]{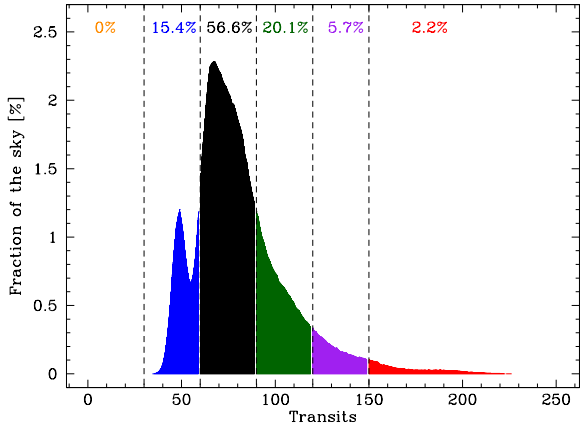}
\caption{Fraction of the sky vs the expected end of mission number of transits.
About 15.4\% of the sky is expected to be observed 30-60 times, $\sim 56.6\%$ of the sky is expected to be observed 60-90 times and $\sim 20.1$\% 90-120
times. The remaining 7.9\% is expected to be observed 120-240 times.
J. de Bruijne, private communication.
} 
\label{fig:transitssky}
\end{figure}

Gaia\footnote{\tt http://sci.esa.int/science-e/www/area/index.cfm?fareaid=26} 
satellite is  constituted by
two  1.45~m~$\times$~0.5~m telescopes pointing in 
two directions separated 
by a  106.5 deg basic angle and merged
 into a single focal plane. 
The satellite will operate from L2, $1.5\times10^6$ km from the Earth.
Launch date is  set  in June 2013, the nominal mission ends after 5 years. 
Gaia will continuously scan the whole sky spinning  at 60''/sec.
Due to the basic angle, each sky field  is observed 
again after 106.5~minutes\footnote{and each 
observation consist on a transit on 9 CCD, with an integration time of 
4.4~s per CCD}  and it is possibly re-observed after a full satellite 
rotation,  4h~13m later.
This time sampling pattern is repeated sequentially several times 
as the result of the slow spinning  and precession\footnote{the 
precessional period is 63 days long} of the satellite.
The same field  may be observed again after $\sim$30 days on average
and $\sim$80 times on average over the 5 years,
but the scanning law is largely uneven and transits may vary 
between 40 and 250 times. Regions lying at
ecliptic latitude $\pm 45\degr$ will be scanned on average more often than other locations.
As shown in Fig.~\ref{fig:transitssky} more than one half of the
sky will be observed 60-90 times. 

Recent estimates of the observable SNe (Type Ia + CC)
based on  a 
 limiting magnitude $G=19$ predicts
$\simeq~6000$ SNe over the 5-year mission and the most distant observed  type Ia SNe
will be at  $\simeq 500Mpc$ ($z\simeq0.12$). About 1/3 of the SNe is expected
to be observed before maximum.

With the aim to obtain a preliminary   characterization of 
SNe 
detected by Gaia, we exploit our simulation tool which 
was developed to estimate the expected number of  both  SNe types, along with their redshift 
and apparent magnitude distributions, for any given SN search. 

There are two types of inputs for this simulation: the transient properties (photometric evolution and rate of occurrence)
and the survey strategy (survey area, limiting
 magnitude and monitoring cadence).

We  considered  "classical" SN classes: Ia, IIP, IIL, IIn and Ib/c. 
For each of these SN types we need to know: a) the template light and color 
curves along with absolute magnitude at maximum and dispersion 
(Li et al. 2011); b) the K-correction as a function of redshift from 
the template to the survey observing band (cf. \citealt{botticella2008}); 
c) the current best estimate of the SN rate evolution with redshift 
(\citealt{botticella2008} and references therein).

SNe occur in galaxies with a rate that depends on the galaxy type and, for 
a given type, is, at least in first approximation, proportional to the 
galaxy luminosity. Therefore, in principle, an accurate simulation would require 
as input the catalog of galaxies monitored by the search along with their 
type and luminosity.
 However, if the volume of the search, which is defined
 by the survey area times its depth, is large enough we can rely on the 
Universe homogeneity on large scale and adopt volumetric rates. For the 
current simulation we assume that this is the case for the Gaia survey that, 
although not very deep, is  all sky.
We derived the transit distributions for the different 
sky pixels from the Gaia  sky coverage map adopting a pixel resolution of 0.84 square
degrees.  As a further simplification we assumed that the transits occurring 
within 6 hours from the first observation (i.e. after a full satellite
rotation)
 see the same sky while more distant transits are evenly 
distributed within the 5 years of the mission. For our aim to estimate 
the total number of SNe detected by Gaia and given the timescale of 
SN evolution this is  quite a good approximation.

\begin{figure}[!t]
\includegraphics[angle=0,scale=.41]{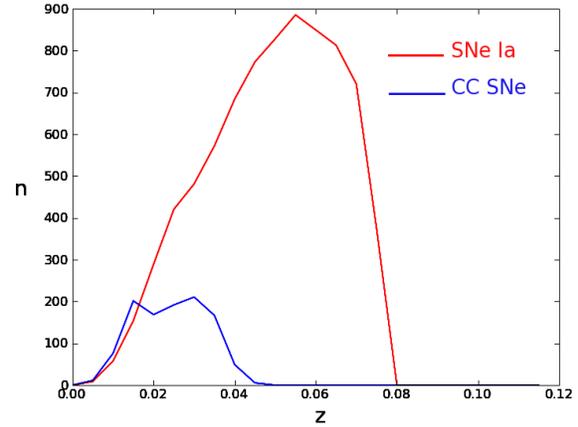}
\caption{Redshift distribution of the expected SNe detected by Gaia.
Red line: SN Ia, blue line: Core Collapse SNe. 
} 
\label{fig:SNzdistribution}
\end{figure}

We expect that Gaia will detect about 6300 SNe, most of them Ia (85\%) and the remaining CC SNe. Of the SNe Ia detected about 30\% will be discovered before 
maximum light. SNe Ia will be discovered up to redshift $z\simeq 0.1$ with a peak 
around $z \simeq 0.06$. Instead almost all CC SNe will be at redshifts 
$z<0.05$ (Fig.~\ref{fig:SNzdistribution}). Based on tests with different 
assumptions, the uncertainty of the simulation is $\sim$10-15\%.
Our results agree very well with the recent estimates by S. Hodgkin (private communication). 
\\
Even if  the expected number of  SNe detected by Gaia
decreased in the most recent estimates (see \citealt{hog1999,belokurov2003}, this work), 
it still represents a significant contribution to the current  discovery rate (Fig.\ref{fig:sndiscoveryrate1}).

\section{Gaia science alerts and ground based observations}\label{sec:gaiaandgroundbasedobservations}

The Gaia Science Alerts stream\footnote{see 
the Gaia Science Alerts Working Group wiki http://www.ast.cam.ac.uk/ioa/research/gsawg}, with almost real time transient events,
will be the first Gaia data released to the astronomical community.
It should be noted that transients will not be proprietary, as is often the case
nowadays in many  wide-field surveys  undertaken by "private" projects.
All transients data  will be
distributed  immediately after the detection in accordance with the ESA agreement.
This will allow the scientific community to investigate
interesting targets immediately.
Gaia  is expected to disseminate several SN alerts per day along the 
five years mission, however many of these SN candidates  will be either 
already discovered by other SN survey or very late due to the satellite time sampling.
Only the  regions with frequently repeated visits will produce SN candidates which are  very young, hence potentially most interesting.
In all large surveys the rejection
of false positive, the confirmation of the SN candidates
and their prompt follow-up represent a challenging issue.
The Gaia time sampling will provide a photometric confirmation and a partial follow up,  since for  each observed object
will  be available  a color and a low resolution  spectrum  
provided by the  two low dispersion
slitless prism spectrographs, the Blue Photometer (BP) and the Red Photometer 
(RP), that cover the 3300-6800\AA\, and 6400-10500\AA\, ranges respectively,
with a spectral resolution varying from 20 to 100 approximately.
In principle, 
these data can be very valuable to obtain a prompt 
transient classification as SN and possibly the SN type.
Each object brighter than magnitude 17 will be also observed through
the Radial Velocity Spectrometer (RVS), providing an high resolution 
($R=11500$) spectrum in a narrow region  (8470-8740\AA) around the calcium 
triplet. Unfortunately most of the expected SNe will be fainter
than magnitude 17.
\\
Due to the irregular time sampling and to the
instrumental setup that is not optimal for SN follow-up,
ground based observations are highly desirable.
A prompt follow up 
is also desirable for studying early phases properly.
The astronomical community, including the Italian one,  is trying to set up
a network to secure a  prompt reaction to the announcement of the transients detected by the satellite (\citealt{burgon2011}) 
and an intensive follow-up of the most interesting transients.
The most distant  SNe Ia accessible to Gaia are at  $z\simeq 0.12$, 
hence $2-4$~m class telescope, or even smaller  for brighter objects, located 
in both the hemispheres, will be suitable.
Follow-ups may be limited to the most intensively
regions monitored by Gaia to achieve the maximum efficiency. 
\\
Some of us are involved in the Public ESO Spectroscopic Survey for Transient
 Objects  (PESSTO) at NTT and in  the Gaia-ESO  Survey (GES) at VLT.
The goal of  PESSTO  is to exploit the large allocations
of telescope time available through an ESO public survey to assemble 
data on all forms of SNe covering the full range of parameter space that 
the current surveys now deliver  (luminosity, host metallicity, explosion mechanisms). The survey is expected to deliver time series spectra of 150 transients
and optical and NIR light curves for tha majority of these transients.
In order to select the targets, $\sim 2000$ SNe are expected to  be classified, 
and all information on these will be publicly available, providing
 an alert list available to the global community.
PESSTO   will start on April 2012 and 90 nights per year over 4 years (with
a possible 1 year extension) have already been granted.
The public spectroscopic survey GES started  on Dec. 31, 2011 
and a total of 240 nights over 4 years (with
a possible 1 year extension)
have already been awarded to the project.
The main goal of the GES is to study the formation and evolution of 
the Milky Way and its stellar populations.
The observation of  $\sim 10^5$ stars, systematically covering all
 major components of
the Milky-Way will provide valuable complementary data to Gaia
and vice versa.

\section{Gaia contribution to SN studies}\label{sec:conclusions}
The  unified study of  SN progenitor stars and SN  explosion characteristics as well as the analysis of the SN rate evolution with redshift  are promising to solve important issues about these intriguing events and  to exploit SNe as cosmological standard candles and as a probe of the SFR.
Searches for transients to increase the sample of both standard and  peculiar  SNe   are  required.
While transients are not the founding scientific rationale of Gaia,  it has become clear that the mission  will be a competitive, all sky transient survey without
any input object bias and will collect a huge data-set of local SNe, with 
a significant fraction of under-luminous and over-luminous events.
A community of interested scientists is being formed and is in contact to discuss the science
opportunities with this mission.
\\
The big and homogeneous samples provided by Gaia,
coupled with  a homogeneous data treatment,
will allow robust statistical studies. Relatively rare/peculiar SNe
or new transients  will be studied in a statistically significant way, 
subclassifying them into  much finer grids if necessary. 
These large statistical studies  will significantly improve our understanding 
of the SN explosions and they  will substantially reduce systematic errors in 
distance estimates, providing tighter constraints on cosmological parameters.
\\
The extremely accurate positions provided by Gaia
will allow a detailed study of the SN spatial distribution
in different galaxy types and environments.
The precise spatial location together with additional data 
from ground-based observations may provide evidence for possible progenitors 
and correlations between SNe properties, location/environment and host type.
Distances and redshifts will also be useful to probe the local velocity field.


\bibliographystyle{spr-mp-nameyear-cnd}

\end{document}